\documentclass[conference]{IEEEtran}
\usepackage[T1]{fontenc}
\PassOptionsToPackage{hyphens}{url}
\usepackage{booktabs,graphicx,url,amsmath,amssymb,xcolor,microtype,array,dblfloatfix}
\usepackage[hidelinks]{hyperref}
\usepackage{flushend}
\newcommand{\LZ}{LZ77}
\newcommand{\head}{HEAD}
\newcommand{\pkgurl}{\url{https://osf.io/5nwgv/?view_only=34532f33e19149829435aa819d476191}}
\newcolumntype{P}[1]{>{\raggedright\arraybackslash}p{#1}}
\begin{document}
\title{Git History Is a Pangenome: Repetition in Code Training Corpora and What Deduplication Removes, Keeps and Misses}
\author{\IEEEauthorblockN{Saurabh Chopra}\IEEEauthorblockA{GitLab Inc.}}
\maketitle

\begin{abstract}
Code models are increasingly trained on the full history of software repositories, every version of every file, rather than on a single snapshot. The tools that deduplicate training data were built for snapshots, and what they do to a history has not been measured. We measured it on a pre-registered random sample of 306 public repositories, 25 mature projects, and the 12 repositories behind the SWE-bench benchmark. A history is mostly repetition: in the vocabulary of text indexing, a pangenome, one text in many near-identical variants. The history of a typical repository holds the equivalent of 19 copies of its current code, and a mature repository holds 89. The reason is that each edit changes a few lines and leaves the rest of the file in place: nine tokens in ten of a changed file already existed before the change. Exact-match deduplication removes none of these copies, since no two versions are identical. Large training pipelines already deduplicate their snapshots by grouping near-identical files and keeping one. Applied across versions, the same step brings the repetition down to about two copies. But two problems remain. The deduplicator keeps one version of each file, and in widely used implementations, including the one we measured, the rule for choosing that version ignores dates, so for half of all files the version kept for training is an old one. Sorting versions newest first before deduplicating fixes the problem at almost no cost. The second problem is benchmark answers. Benchmarks such as SWE-bench test a model on real bug fixes, and a merged fix survives in a median of 18 later versions of the files it changed. The deduplicator cannot be relied on to remove them, because it keeps one version of each group and that version may contain the fix. Therefore, removing benchmark fixes needs a separate step that checks every version. We turn the measurements into a short set of rules for teams that train on code history, and release the instruments and frozen outputs.
\end{abstract}
\begin{IEEEkeywords}
code language models, training data, deduplication, decontamination, mining software repositories
\end{IEEEkeywords}

\section{Introduction}
Consider a file, \texttt{utils.py}, in a project that is five years old. It has been edited 30 times. The repository's history therefore contains 30 versions of \texttt{utils.py}, and each version is about 95\% identical to the one before it. For most of the last decade, code models never saw those 30 versions. Training sets such as The Stack took one snapshot of each repository, the latest, and \texttt{utils.py} appeared once~\cite{kocetkov2022stack,lozhkov2024starcoder2}. That is changing. To teach models how code is changed, and not only how it looks when finished, newer training sets take the history. CommitPack pairs the before and after contents of the file changed by each of 57.7 million commits~\cite{muennighoff2024octopack}. StarCoder trained on a 64~GB commit set~\cite{li2023starcoder}. Seed-Coder ingests about 100 billion tokens of commit data~\cite{seedcoder2025}. Training sets for coding agents replay repositories at thousands of past commits~\cite{pan2024swegym,jain2025r2egym}. In a training set built from history, \texttt{utils.py} appears 30 times, or 60 times when each change is stored as a before-and-after pair. Text indexing has a name for a collection like this, one text in many near-identical variants: a pangenome. A repository history is therefore a pangenome.

The step that is supposed to remove repeated text does not recognize the 30 versions as repeats. Exact deduplication compares file hashes, and no two of the 30 versions have the same hash. The commit-level deduplication that history datasets use compares the change itself, and every one of the 30 edits was a different change~\cite{eliseeva2023commitchronicle,schall2024commitbench}. The one tool that would notice is similarity-based deduplication: MinHash with locality-sensitive hashing (LSH), which groups files whose text overlaps heavily and keeps one file per group. Large training pipelines already run it, but on the snapshot, comparing the single current copy of each file against the files of other repositories, not on the versions of the same file~\cite{allal2023santacoder,li2023starcoder}. Point the deduplicator at the 30 versions and it will group them and keep one. Two questions follow at once. Which of the 30 does it keep? And if version 22 contains a bug fix that a benchmark like SWE-bench later tests, does keeping one version remove the fix, or preserve it?

Code training sets come in three kinds, and the findings below apply to them differently. \emph{Snapshot corpora}, the older kind, hold one version of each repository, the latest; the deduplicator already runs on them, choosing among copies of a file across repositories. \emph{Full-file history corpora} treat each version of a file as a training document and charge loss on the whole of it; a commit dataset trained on whole before-and-after files belongs here. For those the deduplicator and its keep rule (which member of a group it keeps) matter directly. \emph{Edit corpora} train on the change rather than on the whole file, as in StarCoder's change windows or an agent's replay of a commit; for those the intermediate versions are the point, so nothing should be collapsed. The cost of an edit corpus is different: the unchanged nine tenths of a file is emitted twice in every pair, and the known remedy is to train on a window around the change rather than on the whole file~\cite{li2023starcoder}. Benchmark fixes have to be kept out of all three kinds. For each finding below we say which kind it serves.

Two facts about code version histories have been known for twenty years: a history is larger than its snapshot, and most commits are small~\cite{purushothaman2005small,alali2008commit,hindle2008large}. What has not been measured is what follows from those facts for a training corpus: how much of a history the deduplicator removes, which version it keeps, and how many copies of a benchmark fix a history contains. We measured all three on one pre-registered random sample of 306 public, permissively licensed repositories in eight languages, with 25 hand-picked mature repositories as the upper tail and the 12 SWE-bench repositories for the question about benchmarks. The deduplicator we measure is BigCode's released script, run unchanged~\cite{bigcode2023neardedup}.

The paper makes three contributions. First, to our knowledge the first measurement of how much a repository history repeats its current code, on a pre-registered sample, and of how much of that repetition BigCode's released deduplication script, the one used for The Stack, removes. Second, the finding that the script's keep rule discards the current version of half of all files, and that sorting versions newest first corrects it at almost no cost. Third, the finding that a merged benchmark fix is present in a median of 18 later versions of the files it changed within the repository's history, so a team that trains on that history has to remove benchmark content from every version that reaches its corpus; deduplication, which keeps at least one version of each group, does not do it for them. The instruments, frozen outputs, pre-registrations and re-derivation scripts are released. Four questions organize the paper:
\begin{itemize}
\item \textbf{RQ1, Removes.} How much does a history repeat, and what does the deduplicator remove? \emph{19$\times$ the current code in a typical repository and 89$\times$ in a mature one; exact matching removes none; the deduplicator leaves about 2$\times$ in a full-file corpus.} (Full-file and edit corpora; Section~\ref{sec:rq1}.)
\item \textbf{RQ2, Why.} What does a commit add? \emph{8.7\% new bytes in a typical repository and 1.9\% in a mature one; nine tokens in ten already existed.} (Edit corpora; Section~\ref{sec:rq2}.)
\item \textbf{RQ3, Keeps.} When the deduplicator keeps one version, which one? \emph{The current version of only 40--52\% of files; keep-newest (keeping the most recent member of each group) raises it to 94\% in both populations, at almost no cost.} (Full-file and snapshot corpora; Section~\ref{sec:rq3}.)
\item \textbf{RQ4, Misses.} What can the deduplicator not remove? \emph{A benchmark fix in a median of 18 later versions, which the deduplicator cannot be relied on to remove.} (All three kinds; Section~\ref{sec:rq4}.)
\end{itemize}

\section{Background and Related Work}
This work sits at the meeting point of three lines of research. Work on deduplicating code training corpora made similarity-based deduplication a standard step, but applied it to snapshots; the history datasets that followed deduplicate by commit, and to our knowledge no published corpus has run the deduplicator across the versions of a file. Work on repeated training data established that exact repeats waste compute and can degrade a model; it did not measure near-identical versions of one file, which is what a history consists of. Work on benchmark decontamination made substring matching and repository exclusion the standard methods; both were designed for data in which a benchmark fix appears once. The measurements in this paper address those three gaps.

\textbf{Deduplication of snapshot corpora.} The Stack~v1 near-deduplicates with MinHash and LSH at a Jaccard threshold of 0.85 over single-token shingles~\cite{kocetkov2022stack}. SantaCoder moved to 5-token shingles at 0.7, which removed a further 16--20\% of files~\cite{allal2023santacoder}; StarCoder and The Stack~v2 kept that setting~\cite{li2023starcoder,lozhkov2024starcoder2,bigcode2024stackv2card}, implemented with 256 hash permutations~\cite{bigcode2023neardedup}. Lee~et~al.\ showed on web text that deduplication improves language models and that about 7\% of C4 tokens sit inside 50-token repeats~\cite{lee2022deduplicating}; RefinedWeb found that an exact-substring pass removed a further 18\% of what MinHash had left~\cite{penedo2023refinedweb}. Pipelines differ on scope: DeepSeek-Coder deduplicates per repository~\cite{guo2024deepseekcoder}, OpenCoder per file~\cite{huang2024opencoder}. Embedding-based methods such as SemDeDup~\cite{abbas2023semdedup} and D4~\cite{tirumala2023d4} could merge versions that MinHash separates, but they have been applied to web text, not to code histories, and they too act on whole documents.

\textbf{Deduplication of history datasets.} CommitChronicle drops commits whose message or diff hash repeats~\cite{eliseeva2023commitchronicle}. CommitBench removes samples with the same diff~\cite{schall2024commitbench}. OctoPack filters by message and license and notes that ``the same code file may appear multiple times''~\cite{muennighoff2024octopack}. StarCoder's commit data was not near-deduplicated~\cite{li2023starcoder}. None of these datasets asks whether successive versions of a file are near-copies of each other, which is the question this paper measures.

\textbf{Repeated data and its cost.} Hernandez~et~al.\ found that repeating 0.1\% of a corpus 100 times degrades an 800M-parameter model to the loss of a 400M one~\cite{hernandez2022scaling}. Muennighoff~et~al.\ found diminishing returns beyond roughly four repeats of a dataset~\cite{muennighoff2023scaling}. Repeated sequences are memorized at rates that rise superlinearly~\cite{kandpal2022deduplicating} or log-linearly~\cite{carlini2023quantifying} with their count, and Chudnovsky~et~al.\ find that the penalty of repeating a subset peaks at an intermediate count~\cite{internalrepetition2026}. StarCoder's authors trained 80\% of the time on a window of 0--32 lines around each change rather than on the full file, ``to avoid spending too much compute budget on learning to copy the file content''~\cite{li2023starcoder}. All of these results concern exact repeats. Near-copies of a file across its versions have not been measured, and we make no claim about their training effect.

\textbf{Keeping benchmark content out of training data.} StarCoder and StarCoder2 remove training files that contain the solutions or prompts of their evaluation benchmarks by substring matching~\cite{li2023starcoder,lozhkov2024starcoder2}. Training sets built for coding agents avoid SWE-bench~\cite{swebench2024} by excluding its repositories or by cutting off at its dates~\cite{pan2024swegym,jain2025r2egym,yang2025swesmith}. Contamination has also been documented on the benchmark itself: models name the files a SWE-bench Verified issue touches from the issue text alone more often than for issues outside the benchmark~\cite{liang2025illusion}, over 94\% of its issues predate model training cutoffs~\cite{aleithan2024swebenchplus}, and decontaminated successors draw their tasks from issues newer than the models~\cite{badertdinov2025swerebench}. Those studies ask what a model has absorbed; we measure where in a repository's history the fixes sit. Both practices were designed for snapshots, where a benchmark fix can appear in at most one version of each file. In a history the same fix is present in every later version of the file. RQ4 counts those versions.

\textbf{Redundancy and staleness in repositories.} Gabel and Su found code largely redundant at the scale of one to seven lines~\cite{gabel2010uniqueness}; Lopes~et~al.\ found 70\% of GitHub files to be clones~\cite{lopes2017dejavu}; Allamanis showed that duplication inflates code-ML evaluation~\cite{allamanis2019adverse}. Software Heritage stores each distinct file content once across 30.2 billion contents~\cite{dicosmo2023softwareheritage,swh2026counters}; that is exact deduplication, and RQ1 measures how little of a history it removes. Jahanshahi and Mockus showed that 17\% of the files in The Stack~v2 have a newer version in their repository, bug fixes among them~\cite{jahanshahi2025cracks}; RQ3 measures how a deduplicator's keep rule chooses between such versions when it sees all of them. The compressed-indexing literature~\cite{navarro2021indexingparti} compresses pangenomes with, among other methods, the relative Lempel--Ziv parse~\cite{kuruppu2010rlz}. A repository history is a pangenome in exactly this sense, and the byte attributor of Section~\ref{sec:method} is our own implementation of that parse.

\section{Method}\label{sec:method}
We measure three populations with four instruments. The populations are a pre-registered random sample of 306 repositories in eight languages, a set of 25 mature repositories with 613,849 non-merge commits between them (ordinary commits with one parent; merges are not counted, since they introduce no changes of their own), and the 12 SWE-bench repositories with their 2,294 benchmark instances (each instance is one real GitHub issue, the repository as it stood at the time, and the fix the maintainers merged). The instruments are a repeat counter, BigCode's deduplication script, a byte attributor and an overlap test against each benchmark fix. We describe each in the order the results use them.

\subsection{Repositories}
\textbf{Random sample (RQ1--RQ3).} The sampling frame is the GitHub Search dataset (GHS)~\cite{dabic2021ghs}: 122,189 non-fork repositories with at least 10 stars and 200 commits, a permissive license and a commit since October 2024. Their main language is one of Python, JavaScript, TypeScript, Java, Go, Rust, C and C++. We drew 40 repositories per language uniformly (the seed and draw log are in the package); 306 were eligible, 12 being too small and 2 over the clone cap. Estimates are population-weighted medians, with each language weighted by its share of the frame, and 95\% intervals come from a 10,000-draw bootstrap stratified by language; equal-weight medians are given beside them. The 200-commit floor keeps 20--37\% of the GHS-indexed MIT-licensed repositories per language (the one license group with step-wise counts), so the sample describes active repositories, not the long tail of tiny ones.

\textbf{Mature set (upper tail for RQ1--RQ3).} 25 hand-picked repositories with long histories: 12 SWE-bench repositories (Python), 7 infrastructure repositories (Go, Java, Rust, Python) and 6 large TypeScript, JavaScript and Python repositories. Their median repetition sits at the sample's 91st percentile, so they are the upper tail of the same distribution. Two of the measurements were too slow to run on all 25 mature repositories. Deduplication across versions was run on 10 of them and per-commit attribution on 19; the repeat count and the storage baseline cover all 25, and every measurement covers the whole random sample. Four of the largest SWE-bench histories (Django, SymPy, Matplotlib, scikit-learn) were left out of the deduplication subset for runtime; Section~\ref{sec:threats} discusses the effect.

\textbf{SWE-bench repositories (RQ4).} The 12 repositories and 2,294 test instances of SWE-bench~\cite{swebench2024}, of which 2,229 have a gold patch (the maintainers' fix, stored as a diff) that applies cleanly to its base commit (the repository as it stood when the issue was open); the other 65 are left out: 39 whose gold patch does not apply cleanly to the base commit and 26 whose patch adds no substantive line. All repositories in the three sets are public and were cloned with full history in October 2026.

\subsection{Units}
A \emph{version} of a file is one git blob of at most 5~MB with no NUL byte in its first 8~KB. Each path is assigned a stratum (\texttt{source}, \texttt{test}, \texttt{docs}, \texttt{data}, \texttt{vendored}, \texttt{generated}, \texttt{other}) by heuristics on its directory and file name; ``source'' means first-party source. A commit's \emph{post-image} is the full contents of every text file it added or modified. RQ1 counts every non-merge commit reachable from \head, the set of commits that commit datasets ingest; RQ2 samples non-merge commits from the whole history, stratified by committer year. We measure in bytes, which can be counted exactly, and state token-level claims because byte ratios within a repository track token ratios: on 2,733 first-parent commits of 11 mature repositories, the per-commit share of new content recomputed over tokens (StarCoder2, Qwen2.5-Coder and DeepSeek-Coder tokenizers, 16-token windows) has Spearman correlation 0.95--0.96 with the byte-level share and differs from it by a median 0.2 percentage points.

\subsection{Instruments}
\textbf{Repeat counter.} From the output of \texttt{git log} with \texttt{-{}-no-merges -{}-raw} we enumerate, for every path, its distinct versions and sum their actual sizes, so a version re-committed unchanged is counted once. Versions per path are weighted by the size of the path's latest version, so the count answers how many times the typical byte of today's code appears in the history. Paths present at \head\ and paths deleted before it are reported separately. The corpus is summed per path, so a version reachable under two paths counts twice; the deduplicator counts each version once, so its multiples run slightly below the counter's.

\textbf{The deduplicator: BigCode's script.} We run BigCode's released local script (\texttt{minhash\_deduplication.py}~\cite{bigcode2023neardedup}) over every distinct text version in a repository's history, with its hashing, LSH and keep rule unchanged. Its settings are token 5-gram shingles, 256 permutations and one version kept per connected component; we run it at the Jaccard thresholds 0.7 (the Stack~v2 setting), 0.85 and 0.9. The script keeps whichever member of a group came first in its input. We fed versions sorted by hash, so the survivor is effectively random with respect to time; a pipeline that fed versions newest-first would keep the newest. datatrove, Hugging Face's widely used data-processing library, implements the same method separately and also keeps an arbitrary member of each group~\cite{datatrove2024}. The Spark job that produced The Stack~v2 (\texttt{bigcode-v2/intra\_dedup.py}, launched with an optional \texttt{-{}-rank}) ranks survivors by license, stars, forks and then latest revision date. On version input that order would act as keep-newest. The Stack~v2 gave the job one revision per repository, however, so the order never chose between versions. The job needs a Spark cluster and could not be run here. We ran its hashing locally on a 1.28~GB snapshot corpus of 31 repositories (the 19 SWE-bench and infrastructure repositories of the mature set plus 12 other large repositories, listed in the package). It removes 15.6\% of bytes against the local script's 16.5\%, so the two implementations remove similar amounts. Our own implementation of the same method, written before the script was available, keeps more in the median at every threshold and is used only for the per-file version counts in RQ1, which the script does not output.

\textbf{Byte attributor.} Given a reference text $R$ and a new text $N$, we build the suffix array of $R\cdot\$\cdot N$ with \texttt{libsais}~\cite{grebnov2021libsais} and derive longest-previous-factor information~\cite{crochemore2008computing}. We then parse $N$ greedily into \LZ\ factors whose sources lie in $R$ or earlier in $N$, a relative Lempel--Ziv parse~\cite{kuruppu2010rlz,ziv1977universal}. At a factor length $L$, each byte of $N$ is one of three kinds: copied from the reference, repeated from earlier in $N$, or \emph{new} (a literal, or part of a factor shorter than $L$). Copies from the reference are split into \emph{same file} (present in the previous version of the same file) and \emph{other files}. RQ2 parses each commit's post-image against the repository snapshot just before the commit at $L{=}64$, about 16 tokens, the scale of an edited line. The attributor agrees with a brute-force parser on 300 random pairs. In a recorded control that the package re-runs, inserting one 83-byte line into a 23,796-byte file yields exactly 83 new bytes and the rest same-file. A development-time check, whose log was not retained, attributed 1,495 commits against an unrelated same-language repository and found 0.5\% of source bytes copyable at $L{=}64$, so ``new'' means new to the commit's own repository.

\textbf{Overlap with benchmark fixes.} This instrument is an overlap analysis, the standard test of the decontamination literature, run over the genealogy of each fix: the sequence of later versions of the files it changed. For every SWE-bench instance and every file its gold patch touches, we enumerate every version of that file dated after the instance's base commit. We do so on the first-parent line (the main line of development) and on every ref (every branch and tag). A version \emph{carries the fix} if at least 90\% of the patch's substantive added lines (12 or more non-space characters) occur in it verbatim, indentation preserved. Versions byte-identical to the pre-fix or the post-fix file are set aside and counted separately, so the carrier count covers versions that differ from the fixed file. For each carrier we record an approximate whole-file similarity to the post-fix file: how much of the fixed file's text the carrier still contains, estimated with the byte attributor. The similarity is a byte-level proxy for, not a computation of, the shingle overlap the deduplicator estimates. For patches with at most two substantive lines (23\% of instances) the carrier test is weak: such lines occur before the base commit in 9.1\% of cases, against 2.8\% for longer patches.

\textbf{Git's own storage.} As a baseline we sum, for every version reachable from \head, its delta-compressed size on disk (\texttt{git cat-file -{}-batch-check} with \texttt{objectsize:disk}) and divide by the bytes of \head\ text.

\subsection{Pre-registration and freezing}
The random sample, its estimators and the RQ1 hypotheses were pre-registered (one logged redraw, one logged instrument correction). The attribution instrument and the path strata used by RQ2 were fixed in a written plan with a dated deviations log. RQ2's per-commit sampling design and both of its frames, and RQ4's overlap test (the 90\% line criterion, the 12-character floor and both version frames), were fixed in code but not pre-registered. Outputs were frozen before the text was written, and every departure from the pre-registered design is logged, such as the choice of BigCode's released script over our own implementation as the instrument. Every headline number was re-derived from the raw outputs by a second, independently written script.

\section{RQ1, Removes: How Much Does a History Repeat, and What Does the Deduplicator Remove?}\label{sec:rq1}
\begin{figure*}[!t]\centering\includegraphics[width=0.96\textwidth]{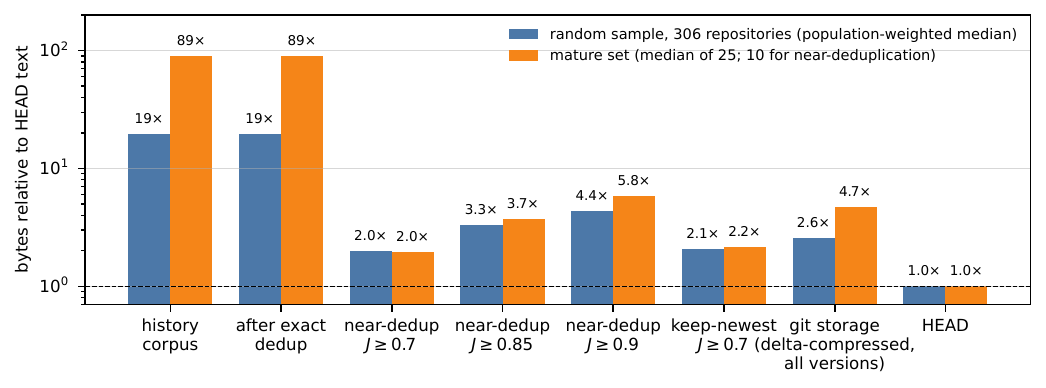}
\caption{Bytes relative to the current code (\head\ text) along the pipeline, for the random sample and the mature set: the history corpus, what survives exact and near-duplicate deduplication at three thresholds, the keep-newest alternative, and git's own delta-compressed storage of every version. The mature set's history and git-storage bars are medians over all 25 repositories; its near-deduplication and keep-newest bars come from the 10 repositories that could be deduplicated across all versions (Section~\ref{sec:threats}).}\label{fig:pipe}\end{figure*}
\begin{table*}[!t]\centering\caption{RQ1. The repeat spectrum of history corpora: all non-merge commits reachable from \head, text paths only. Versions per file are byte-weighted medians, rounded to integers, for all files and for first-party source files; corpus bytes are the sum of every path's distinct post-images (an (old, new) pair corpus is twice that).}\label{tab:rq1}\scriptsize\setlength{\tabcolsep}{3pt}
\begin{tabular}{@{}lrrrrrrr@{}}\toprule
Repositories & Non-merge commits & \head\ text (MB) & \multicolumn{2}{c}{Versions per file} & \multicolumn{2}{c}{Corpus / \head\ bytes} & Deleted-path bytes \\\cmidrule(lr){4-5}\cmidrule(lr){6-7}
 & (median) & (median) & all files & source files & \head\ paths & all paths & \\\midrule
Python ($n{=}37$) & 480 & 1.3 & 5 & 8 & 11$\times$ & 16$\times$ & 20\% \\
JavaScript ($n{=}38$) & 470 & 1.5 & 8 & 6 & 14$\times$ & 29$\times$ & 28\% \\
TypeScript ($n{=}40$) & 473 & 1.4 & 8 & 6 & 16$\times$ & 35$\times$ & 28\% \\
Java ($n{=}39$) & 494 & 2.2 & 3 & 4 & 8$\times$ & 13$\times$ & 35\% \\
Go ($n{=}37$) & 488 & 1.3 & 7 & 8 & 10$\times$ & 15$\times$ & 7\% \\
Rust ($n{=}40$) & 464 & 0.8 & 9 & 12 & 18$\times$ & 23$\times$ & 16\% \\
C ($n{=}36$) & 357 & 4.5 & 2 & 3 & 9$\times$ & 12$\times$ & 13\% \\
C++ ($n{=}39$) & 476 & 4.5 & 4 & 5 & 8$\times$ & 12$\times$ & 19\% \\
\midrule Random, pop.-weighted ($n{=}306$) & 475 & 1.6 & 5 & 7 & 13$\times$ & 19$\times$ & 23\% \\
Random, equal-weighted ($n{=}306$) & 475 & 1.8 & 5 & 6 & 12$\times$ & 18$\times$ & 22\% \\
Mature, median of 25 & 13,224 & 19.0 & 12 & 26 & 53$\times$ & 89$\times$ & 31\% \\
\bottomrule\end{tabular}
\end{table*}

We start with the size of the problem.

\textbf{Finding.} \emph{The history of a typical repository holds the equivalent of 19 copies of its current code and a mature repository 89. Exact deduplication removes none of them. BigCode's deduplicator, run across versions, removes most of them and leaves about two copies (Fig.~\ref{fig:pipe}).}

\textbf{How much repeats.} Table~\ref{tab:rq1} gives the spectrum. In the random sample, the corpus of every version of every file is a population-weighted median 19$\times$ the current code (95\% CI 17--23), and 13$\times$ counting only files still present at \head. The median file at \head\ has 5 versions and the median source file 7. Across the 25 mature repositories (613,849 non-merge commits) the corpus is 89$\times$ the current code (range 18--175$\times$) and 53$\times$ on files still present. The median file has 12 versions and the median source file 26 (46 once renames are followed). Long single-file histories sit at the top: xarray at 175$\times$ with a median of 77 versions per source file, pytest at 145$\times$ with 139. A pre-registered model that predicts a history's size from its commit count and current size fits the mature set well (median prediction 1.04$\times$ the measured size) but predicts an individual repository within a factor of two only 58\% of the time, below the pre-registered 70\%. The repetition is a property of each history, not of its size alone.

\textbf{Is the full history a corpus anyone builds?} Not as such. It is the largest corpus a team could build from a history, and every published recipe is a subset of it. On the mature set, single-file commits alone hold 17.5$\times$ the current code (35$\times$ as (old, new) pairs), commits touching at most five text files hold 58$\times$, and each commit's added lines alone hold 4.9$\times$. The recipes shrink the repetition without removing it. We did not measure the recipes on the random sample. Pairs of before and after contents double the full-file multiple, and deduplication across versions does not change the repeated context inside a pair; Section~\ref{sec:rq2} returns to that cost.

\textbf{What exact deduplication removes.} Nothing beyond a few cross-path duplicates. Every version is a distinct file content, so a filter that compares hashes sees 19 or 89 distinct documents where a reader sees one file.

\textbf{What the deduplicator removes.} No history dataset we know of runs the deduplicator across versions, so Table~\ref{tab:dedup} measures what a team would gain by doing so with the tool it already has. At the default threshold of 0.7 the script keeps 11\% of the sample's history bytes and 2\% of the mature set's; it does its job on most of the repetition. What survives is still 2.0$\times$ the current code in both populations (sample 95\% CI 1.7--2.2; mature range 0.9--12.7$\times$), 3.3--3.7$\times$ at 0.85, and 4.4--5.8$\times$ at the 0.9 threshold some pipelines use. Our own implementation, the instrument the pre-registration named, gives 3.1$\times$ (CI 2.7--3.4) on the sample, above the 1.3--2.5$\times$ its addendum anticipated; we report BigCode's released script as the headline because it is what pipelines run, and our implementation as an upper bound. Typical repositories have fewer versions per file, so less of their corpus is removed, yet the surviving multiple is the same. The survivors are the versions separated by edits large enough to push the 5-gram overlap below the threshold. Under our own implementation, which reports which versions survive, the median source file in the 10-repository mature subset goes from 44 versions to 2 survivors at 0.7 and 7--8 at 0.9; in the random sample from 7 to 1. The deduplicator also misses repetition below the file level. On a deduplicated snapshot corpus of the 306 repositories, a median 4.5\% of a repository's source still lies inside copies of at least 128 bytes of its other files (95\% CI 3.8--5.1\%). That share is a lower bound, and only a substring-level pass of the ExactSubstr kind~\cite{lee2022deduplicating,penedo2023refinedweb} would reach it.

\textbf{What the repetition costs.} Without the deduplicator, a training run over the full history reads the typical byte of today's code about 7 times in a typical repository and 26 times in a mature one (the median versions per source file; the 19$\times$ and 89$\times$ corpus multiples are byte totals dominated by heavily edited files and deleted paths). Each reading comes inside a file that is 77--93\% unchanged (Section~\ref{sec:rq2}). After the deduplicator, about twice. The effect on model quality is a bound taken from the repeated-data literature, not a measurement of ours. If near-copies behave like exact repeats, the data-constrained scaling law of Muennighoff~et~al.~\cite{muennighoff2023scaling} makes a file seen 7 times worth about 6 fresh copies, and one seen 26 times about 13.

\textbf{Git already stores it.} The repetition is cheap to store and expensive only to train on. Git keeps every version reachable from \head, delta-compressed, in a median 2.6$\times$ the current code in the random sample (1.4$\times$ the whole \head\ tree) and 4.7$\times$ in the mature set (4.1$\times$ the tree; 78$\times$ uncompressed). Stored losslessly, a history costs a few snapshots' worth of bytes, about as much as the deduplicator leaves after discarding versions (Fig.~\ref{fig:pipe}).

\textbf{For the corpus builder.} A team building a full-file history corpus should run the deduplicator across versions, not only within the snapshot.

\begin{table*}[!t]\centering\caption{RQ1 and RQ3. BigCode's released local script applied to every distinct text version of a repository's history, one version kept per connected component. ``Kept share of corpus'' is the share of history bytes that survive. ``Current version kept'' is the share of files at \head\ whose current version survives under the script's rule (first in input, here the smallest hash); ``keep-newest'' keeps the most recent member of each component instead. Language rows are medians over repositories. Mature set: 10 repositories (291k versions; eight SWE-bench repositories and two infrastructure repositories, listed in the package).}\label{tab:dedup}\scriptsize\setlength{\tabcolsep}{3pt}
\begin{tabular}{@{}lrrrrrr@{}}\toprule
Repositories & \multicolumn{3}{c}{Kept bytes / \head, smallest-hash rule} & Kept share of corpus & Current version kept & Keep-newest, kept / \head \\\cmidrule(lr){2-4}
 & $J{\ge}0.7$ & $J{\ge}0.85$ & $J{\ge}0.9$ & ($J{\ge}0.7$) & ($J{\ge}0.7$) & ($J{\ge}0.7$) \\\midrule
Python ($n{=}37$) & 2.3$\times$ & 3.2$\times$ & 3.7$\times$ & 15\% & 54\% & 2.4$\times$ \\
JavaScript ($n{=}38$) & 3.1$\times$ & 4.4$\times$ & 6.3$\times$ & 14\% & 50\% & 3.0$\times$ \\
TypeScript ($n{=}40$) & 2.7$\times$ & 5.0$\times$ & 6.3$\times$ & 8\% & 55\% & 2.8$\times$ \\
Java ($n{=}39$) & 1.5$\times$ & 2.4$\times$ & 3.1$\times$ & 14\% & 45\% & 1.6$\times$ \\
Go ($n{=}37$) & 1.8$\times$ & 2.9$\times$ & 4.3$\times$ & 13\% & 53\% & 2.0$\times$ \\
Rust ($n{=}40$) & 1.7$\times$ & 3.0$\times$ & 4.5$\times$ & 8\% & 52\% & 1.9$\times$ \\
C ($n{=}36$) & 1.3$\times$ & 2.1$\times$ & 2.5$\times$ & 12\% & 52\% & 1.4$\times$ \\
C++ ($n{=}39$) & 1.5$\times$ & 2.1$\times$ & 2.9$\times$ & 18\% & 44\% & 1.5$\times$ \\
\midrule Random, pop.-weighted ($n{=}306$) & 2.0$\times$ & 3.3$\times$ & 4.4$\times$ & 11\% & 52\% & 2.1$\times$ \\
Random, equal-weighted ($n{=}306$) & 1.8$\times$ & 3.0$\times$ & 4.1$\times$ & 11\% & 51\% & 1.9$\times$ \\
Mature, median of 10 & 2.0$\times$ & 3.7$\times$ & 5.8$\times$ & 2\% & 40\% & 2.2$\times$ \\
\bottomrule\end{tabular}
\end{table*}
\section{RQ2, Why: What Does a Commit Add?}\label{sec:rq2}
\begin{table*}[!t]\centering\caption{RQ2. Per-commit attribution of each commit's post-image against the repository snapshot before the commit, up to 300 non-merge commits per repository drawn from the whole history and stratified by committer year. ``Commits'': median measured commits per repository. ``Parent $<$1~MB'': share of sampled commits whose parent snapshot holds under 1~MB of text. ``New'': pooled share of post-image bytes in literals and factors shorter than $L$. ``Median commit'' and ``Commits $<$1\%'': the per-commit distribution of that share. ``Gini'' and ``Top-10\%'': concentration of new bytes across commits. ``Same file'': share already present in the previous version of the same file. Random-sample medians cover the 305 repositories with at least 20 measured commits (one C++ repository is excluded, Section~\ref{sec:threats}).}\label{tab:commit}\scriptsize\setlength{\tabcolsep}{2pt}
\begin{tabular}{@{}lrrrrrrrrr@{}}\toprule
Repositories & Commits & Parent $<$1~MB & \multicolumn{2}{c}{New, pooled} & Median commit & Commits $<$1\% & Gini & Top-10\% & Same file \\\cmidrule(lr){4-5}
 & (median per repo) & & $L{=}64$ & $L{=}32$ & ($L{=}64$) & ($L{=}64$) & & & ($L{=}32$) \\\midrule
Python ($n{=}37$) & 295 & 57\% & 11.5\% & 8.4\% & 3.4\% & 27\% & 0.87 & 81\% & 82.3\% \\
JavaScript ($n{=}38$) & 293 & 69\% & 8.2\% & 5.0\% & 2.8\% & 32\% & 0.91 & 88\% & 84.0\% \\
TypeScript ($n{=}40$) & 296 & 48\% & 6.3\% & 3.9\% & 3.4\% & 31\% & 0.84 & 74\% & 90.7\% \\
Java ($n{=}39$) & 294 & 24\% & 8.9\% & 5.1\% & 2.4\% & 35\% & 0.85 & 77\% & 79.0\% \\
Go ($n{=}37$) & 296 & 65\% & 8.9\% & 7.2\% & 3.2\% & 28\% & 0.82 & 74\% & 80.1\% \\
Rust ($n{=}40$) & 298 & 92\% & 6.3\% & 4.0\% & 2.8\% & 30\% & 0.79 & 66\% & 91.3\% \\
C ($n{=}36$) & 292 & 21\% & 10.3\% & 7.6\% & 3.3\% & 27\% & 0.87 & 82\% & 83.4\% \\
C++ ($n{=}38$) & 291 & 26\% & 10.6\% & 6.9\% & 2.9\% & 29\% & 0.87 & 80\% & 84.3\% \\
\midrule Random, pop.-weighted ($n{=}305$) & 295 & 54\% & 8.7\% & 5.6\% & 3.2\% & 30\% & 0.85 & 78\% & 87.0\% \\
Random, equal-weighted ($n{=}305$) & 295 & 51\% & 8.3\% & 5.5\% & 3.0\% & 30\% & 0.85 & 78\% & 86.4\% \\
Mature, median of 19 & 298 & 2\% & 1.9\% & 1.3\% & 1.0\% & 50\% & 0.88 & 83\% & 95.1\% \\
\bottomrule\end{tabular}
\end{table*}
\begin{figure}[!t]\centering\includegraphics[width=\columnwidth]{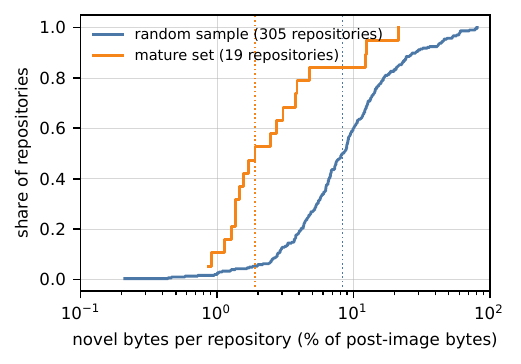}
\caption{Cumulative share of repositories by new-byte share (pooled new bytes over post-image bytes at $L{=}64$), for the random sample (305 repositories) and the mature set (19); dotted lines mark the medians.}\label{fig:nov}\end{figure}

The deduplicator removes so much because each version differs so little from the last. This section measures how little.

\textbf{Finding.} \emph{In the typical repository, 8.7\% of a changed file's bytes are new; in a mature repository, 1.9\%. The rest already existed in the repository before the change, mostly in the same file, and a tenth of the commits carry about 80\% of the new bytes.}

\textbf{How much a commit adds.} We attributed 83,924 sampled commits of the 305 sample repositories that enter the medians and 5,645 commits of 19 mature repositories (Table~\ref{tab:commit}, Fig.~\ref{fig:nov}). In the median repository of the sample, 8.7\% of a changed file's bytes are new at $L{=}64$ (95\% CI 7.2--9.3\%) and 87.1\% already occur somewhere in the snapshot before the commit (the remaining 4\% repeats text that appears earlier in the same commit's changed files). In the mature set, 1.9\% are new and 95.8\% already occur. The median commit adds 3.2\% in the sample and 1.0\% in the mature set. Read the other way, nine tokens in ten of a changed file are copies of bytes the repository already contained.

\textbf{Mostly in the same file.} Split by file, 77\% of the sample's post-image bytes (pooled over 80,225 modifying commits; 87\% for the median repository) and 93\% of the mature set's (pooled; 95\% for the median repository, Table~\ref{tab:commit}) were already in the previous version of that very file. For the median commit the share is 97\% and 99\%. An (old, new) pair of the CommitPack kind emits that unchanged context twice per training example.

\textbf{Concentrated in a few commits.} New bytes are concentrated in both populations. In the sample and the mature set respectively, the Gini coefficient of new bytes per commit is 0.85 and 0.88, and the top 10\% of commits carry 78\% and 83\% of the new bytes. The single commit with the most new bytes carries 17\% and 33\% of them. The three mature repositories above 5\% new bytes owe it to non-code content: their single most novel commit, a vendored dependency tree (twice) or re-run example notebooks, carries 37\%, 69\% and 37\% of their new bytes. Restricted to code-only commits (18 of the 19 mature repositories have them), the mature set shows 1.2\% new bytes, a Gini of 0.73 and a top-10\% share of 58\%.

\textbf{Mature repositories add less.} Typical repositories are small and young. In the median repository of the sample, 54\% of commits have a parent snapshot under 1~MB, against 2\% in the mature set, so an ordinary commit is a larger fraction of the tree. Size is not the whole story. Across equal-count terciles of the sample by non-merge commits, the equal-weight medians fall from 9.9\% to 9.0\% to 6.4\%, and the 25 sample repositories with at least 2,300 non-merge commits sit at 3.8\%. The structure is the same everywhere: most bytes already present, the new ones concentrated in a few commits. Two robustness checks hold: sampling only first-parent commits gives 2.2\% on the mature set with a top-10\% share of 77\%, and attributing merge commits against their first parent gives 3.0\% with the same concentration.

\textbf{For the corpus builder.} For edit corpora the finding points away from deduplication. The versions are the signal, and the arithmetic is direct: nine tokens in ten of a typical training example are text the model is asked to reproduce unchanged, so a model that is charged loss on the whole file spends most of its compute learning to copy. Whether that harms quality is not measured here; what is measured is the share. Training on a window around the change, as StarCoder did~\cite{li2023starcoder}, already avoids most of that cost. A one-line fix is low in new bytes and is exactly what an agent must learn, so the lever is where loss is charged, not which commits are sampled.

\section{RQ3, Keeps: When the Deduplicator Keeps One Version, Which One?}\label{sec:rq3}
A team that follows RQ1 and deduplicates across versions hands the deduplicator a decision: of the 30 versions of \texttt{utils.py}, which one stays? The same decision already governs snapshot corpora, where the deduplicator chooses among copies of a file across repositories.

\textbf{Finding.} \emph{BigCode's deduplicator keeps the current version of only 40\% of files in the mature set and 52\% in the random sample. Keeping the newest version of each group instead retains the current version of 94\% of files in the mature set and costs 0.1--0.2$\times$ the current code in extra bytes.}

\textbf{Which version survives.} The script keeps whichever member of a group came first in its input, in our runs the smallest hash. Its survivors include the current version of only 40\% of files in the mature set and 52\% in the random sample (Table~\ref{tab:dedup}, last columns). For roughly half of all files, then, the version a pipeline on this rule would train on is not the current one: an older version of the file, or a near-copy from another path. Keeping the newest member of each group instead needs only a commit date per version and costs almost nothing: 2.1$\times$ the current code in the random sample against 2.0$\times$, and 2.2$\times$ in the mature set against 2.0$\times$. It retains the current version of 94\% of files in both populations.

\textbf{Does the choice matter?} We did not train models to find out, so the following is argument, not measurement. An older survivor lacks every later change to its file, and RQ4 shows what later changes contain. The fixes that benchmarks are built from spread into a median of 18 later versions, so a survivor chosen without regard to time may be a version from before a fix; how often, we did not measure. Jahanshahi and Mockus found that 17\% of the files in The Stack~v2 have a newer version in their repository, and that about one in six of those newer versions is a bug fix~\cite{jahanshahi2025cracks}. Theirs is a snapshot statistic and ours a deduplication statistic, but the mechanism is the same. When the deduplicator sees every version, the share of files left with a stale survivor is not 17\% but about half. And current versions match current APIs and conventions, which is what a model asked to write code today is judged on. On those grounds the newest version is the better default, and the measurement shows that choosing it is nearly free.

\textbf{For the corpus builder.} A full-file history corpus built on BigCode's local script, on datatrove, or on any rule that keeps the first version seen should sort versions newest first before deduplicating; the measurement above is the gain. Snapshot corpora face the same rule when the deduplicator groups copies of a file across repositories, and would gain by the same mechanism, though we did not measure that case.

\section{RQ4, Misses: What Can the Deduplicator Not Remove?}\label{sec:rq4}
\begin{table*}[!t]\centering\caption{RQ4. SWE-bench gold fixes across the versions of their patched files, per repository. A version carries the fix if at least 90\% of the patch's substantive added lines occur in it verbatim. ``Instances (measured)'': SWE-bench instances and, in parentheses, those whose gold patch applies cleanly. ``Later versions'': first-parent versions dated after the base commit that differ from the fixed file. ``Per patched file'': the same count divided by the files the patch touches. ``Instances with $\ge$10'': share of measured instances whose fix recurs in at least ten later versions. ``Carriers $\ge$80\% similar'': share of carrying versions whose approximate whole-file similarity to the fixed file (Section~\ref{sec:method}) is at least 0.8, a proxy for the range the deduplicator groups at Jaccard 0.7. ``Byte-identical copies'': later versions identical to the fixed file, in practice the fixed file itself landing on the main line, counted separately; they are all that exact deduplication would remove.}\label{tab:rq4}\scriptsize\setlength{\tabcolsep}{2pt}
\begin{tabular}{@{}lrrrrrrr@{}}\toprule
Repository & Instances (measured) & \multicolumn{2}{c}{Later versions carrying the fix} & Per patched file & Instances with $\ge$10 & Carriers $\ge$80\% similar & Byte-identical copies \\\cmidrule(lr){3-4}
 & & median & mean & median & & & mean per instance \\\midrule
django/django & 850 (843) & 10 & 26.8 & 8 & 52\% & 81\% & 1.5 \\
sympy/sympy & 386 (372) & 30.5 & 55.1 & 20 & 74\% & 72\% & 1.5 \\
scikit-learn/scikit-learn & 229 (218) & 17 & 30.2 & 11 & 67\% & 75\% & 1.2 \\
sphinx-doc/sphinx & 187 (184) & 22.5 & 41.0 & 16 & 70\% & 73\% & 0.8 \\
matplotlib/matplotlib & 184 (178) & 44 & 70.7 & 29 & 80\% & 85\% & 1.0 \\
pytest-dev/pytest & 119 (111) & 21 & 33.0 & 14 & 65\% & 66\% & 1.1 \\
pydata/xarray & 110 (103) & 42 & 104.1 & 15 & 83\% & 60\% & 1.8 \\
astropy/astropy & 95 (89) & 10 & 19.1 & 6.5 & 52\% & 90\% & 1.3 \\
pylint-dev/pylint & 57 (56) & 19 & 24.8 & 10 & 70\% & 84\% & 1.6 \\
psf/requests & 44 (43) & 28 & 46.3 & 24.5 & 67\% & 61\% & 0.8 \\
mwaskom/seaborn & 22 (21) & 13 & 15.2 & 7 & 67\% & 95\% & 1.5 \\
pallets/flask & 11 (11) & 15 & 16.3 & 5 & 55\% & 68\% & 1.3 \\
\midrule All instances (pooled) & 2,294 (2,229) & 18 & 40.3 & 12 & 64\% & 77\% & 1.4 \\
Median over repositories & & 20 & & 12.5 & 67\% & & \\
\bottomrule\end{tabular}
\end{table*}
\begin{figure}[!t]\centering\includegraphics[width=\columnwidth]{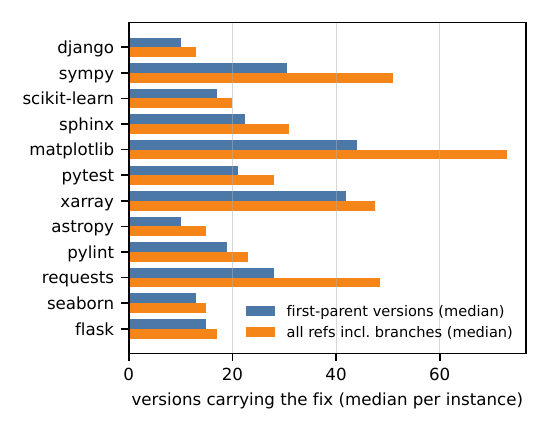}
\caption{Later versions carrying a SWE-bench gold fix, median per instance and repository, on the first-parent line and across all refs. The all-refs counts for Django, SymPy, xarray and Flask are lower bounds: their clones expose only four remote branches.}\label{fig:contam}\end{figure}

Everything a history repeats, it repeats for every file, including the files that benchmarks are built from. Version 22 of \texttt{utils.py}, the bug fix, is also in versions 23 through 30. Two situations matter. In a history corpus built the way they are built today, with no deduplication across versions, all of those later versions are in the training data, and deleting the one file or commit that matches the benchmark leaves the rest. In a corpus that follows RQ1 and RQ3, the deduplicator collapses the versions to one or two per file and keep-newest makes the survivor the newest version, which contains the fix. Deduplication reduces the number of copies; it does not remove the fix. This section measures the first situation and reasons about the second.

\textbf{Finding.} \emph{A merged SWE-bench fix reappears in a median of 18 later versions of the patched files on the main branch, and in 24 across all branches. Removing the fixed file by hash removes one version. The deduplicator keeps one version per group; under keep-newest that version carries the fix, and under the default rule it may. Benchmark content therefore has to be removed in a separate pass over every version that reaches the corpus, on every branch.}

\textbf{What we measured.} Across the 2,229 SWE-bench instances whose gold patch applies, the fix's substantive added lines recur verbatim in a median of 18 later first-parent versions of the patched files that differ from the fixed file (mean 40; per patched file, median 12). 88\% of fixes recur in at least two later versions and 64\% in at least ten (Table~\ref{tab:rq4}, Fig.~\ref{fig:contam}). Within one year of the fix the median is already 8. Per-repository medians run from 10 (Django, Astropy) to 44 (Matplotlib). Branches add more. Across all refs, release branches and backports included (2,268 instances on this pass, which searches every version for the patch's lines instead of applying the patch, so only the 26 patches without a substantive line are excluded), a fix is in a median of 24 distinct versions (mean 61), 1.2$\times$ the main-line count for the median instance. For half of the instances the fix is the latest version on at least two branches. The fixed file itself lands on the main line about once per patched file (a mean of 1.4 byte-identical versions per instance); that one version is all that exact deduplication, or excluding the fixed file by hash, removes.

\textbf{What follows from how the deduplicator works.} We did not trace the benchmark fixes through the deduplicator's groups, so this paragraph reasons from its rule. The deduplicator keeps one version from every group, never none. Under keep-newest, the survivor of a group that contains fixed versions is its newest member, a fixed version unless later edits rewrote the fix's lines, so the fix stays. Under the default rule, a group that mixes pre-fix and post-fix versions may by chance keep a pre-fix one. That is the tension between RQ3 and RQ4, and the filter's place in the pipeline decides how it is settled. How many groups the 18 copies form depends on how far they have drifted. By our approximate similarity, 77\% of carriers are at least 80\% similar to the fixed file and would fall into its group; the rest (median 0 per instance, mean 16) would form further groups, each with its own surviving copy. The similarity is our approximation, not the deduplicator's fingerprint (Section~\ref{sec:threats}), so these are expectations, not measurements.

\textbf{What current practice would do.} A substring or $n$-gram filter run against the gold patches catches every carrier by construction, because every carrier contains at least 90\% of the fix's lines verbatim. The filter has to see every version that reaches the corpus, on every branch, not just the snapshot or the fix commit. It can run before or after deduplication. The fix is removed either way; what differs is the cost and what is left of the file. After deduplication with keep-newest, the filter scans far less, about 2$\times$ \head\ instead of 19--89$\times$, but deleting the surviving copy of a fixed file leaves at most some older, dissimilar versions of it. Before deduplication, the filter deletes every fixed version and keep-newest then keeps the newest unfixed one: roughly the state the benchmark starts from, but stale. A date cutoff set before the fix works on a history as it does on a snapshot. A repository blocklist removes only the copies it can name; forks, vendored copies and mirrors of the 12 repositories were not counted here, so the multiplicities above are lower bounds for a corpus that includes them.

\textbf{Caveats.} For the 23\% of patches with at most two substantive lines the carrier test is weak, as the false-positive floor in Section~\ref{sec:method} shows; the per-instance medians are driven by the longer patches. Django supplies 37\% of SWE-bench instances; medians over repositories are reported beside the pooled ones in Table~\ref{tab:rq4}. We make no claim about the effect of the repeated fixes on benchmark scores and did not measure one; the finding is about what a filter must remove and how many places it must look.

\textbf{For the corpus builder.} Remove benchmark content in its own pass over every version that reaches the corpus, on every branch; whether that pass runs before or after deduplication decides what each benchmark file is left with. All three kinds of corpus need the pass.

\section{Guidance for Training AI Models on Code History}
Each rule below carries the number this paper measured for it and says which kind of corpus it applies to.
\begin{itemize}
\item \textbf{Every pipeline: remove benchmark content in a pass over every version that reaches the corpus, on every branch.} A benchmark fix lives in a median of 18 later versions of the files it changed on the main branch and 24 across branches. Hashing the fixed file removes one of them, a repository blocklist removes only the copies it can name, and deduplication keeps a survivor from every group, which in a group holding the fix is a carrier under keep-newest (RQ4). In a deduplicated corpus, the pass is cheaper after deduplication but may leave only older, dissimilar versions of the file, or none; run before it, it leaves a stale pre-fix version.
\item \textbf{Full-file history corpora: deduplicate across versions.} The deduplicator takes a history from 19--89$\times$ the current code to about 2$\times$ at the default threshold and 4--6$\times$ at the strictest (RQ1). The pass costs compute at corpus scale, but it is a pass the pipeline already knows how to run.
\item \textbf{Full-file and snapshot corpora: keep the newest member of each group.} The time-blind rule of the local script, datatrove and first-seen reimplementations leaves a stale version for half of all files. Keep-newest needs only a commit date per version, retains 94\% of current versions in the mature set, and costs 0.1--0.2$\times$ the current code (RQ3).
\item \textbf{Edit corpora: keep every version; spend the training on the change, not on the rest of the file.} In a before-and-after pair, nine tenths of the file did not change (RQ2). Computing the training loss only on the changed lines stops the model from being taught to copy the rest, but the rest is still read, and reading costs compute; StarCoder instead trains on a window around the change~\cite{li2023starcoder}. A cheaper layout stores the current snapshot once plus only the lines each commit added: 5.9$\times$ the current code instead of 89$\times$, but it drops the lines a commit removed and the surrounding code (RQ1).
\item \textbf{Every pipeline: keep a count of versions per file.} One pass over \texttt{git log -{}-raw} and version sizes, with the released counter and no diffing, gives the versions-per-file spectrum and hence how many times each byte is re-read (RQ1).
\end{itemize}

The first three rules are one observation applied three times. In a history, each file exists in many near-identical versions, and the deduplication tools treat each version as a separate document. The second rule removes the surplus versions, the third decides which version remains, and the first removes the versions that must not remain at all. Together they leave a history corpus that is smaller, free of benchmark content, and current outside the files the benchmark fixes touched. Without the first, they leave one that is smaller and current and still contains the benchmark fixes.

\section{Threats to Validity}\label{sec:threats}
\textbf{What is not measured.} No training run is reported, so the cost of repetition is a bound from literature on exact repeats. How often the deduplicator's time-blind survivor predates a benchmark fix was not measured. We did not run the benchmark filter in either position; what each position leaves behind is deduced from the deduplicator's keep rule, not measured. The recipe variants were measured on the mature set only. The whole-file similarity used in RQ4 is a byte-level proxy that was not calibrated against the deduplicator's fingerprints. The Stack~v2's Spark job was not run end to end; its hashing was. Forks, vendored copies and mirrors of the SWE-bench repositories were not counted. Any effect of the repeated fixes on benchmark scores was not measured.

\textbf{Construct validity.} \LZ\ novelty is a compression notion, not semantic value, and $L{=}64$ is one choice; the package reports $L{=}32$ and $128$ and both pooled and per-commit medians, because large move and vendoring commits dominate pooled sums. Our own implementation of the method is not The Stack~v2's code, so every headline number comes from BigCode's released script, and our implementation is reported only as an upper bound. The overlap test for benchmark fixes is line-based and over-counts for short patches, as reported.

\textbf{Internal validity.} The 10 repositories of the mature deduplication omit Django, SymPy, Matplotlib and scikit-learn, four of the largest histories, which were too large to deduplicate across all versions with our resources. Three of them (SymPy, Django, scikit-learn) have more versions per source file than the median included repository and one (Matplotlib) fewer. Versions per file does not predict the kept multiple (Section~\ref{sec:rq1}), so the direction of any bias is unknown; the random sample carries the headline. Strata come from path heuristics; the largest near-duplicate group in the sample is machine-generated API classes that the path heuristics label as source. 25 sample repositories were measured by the deduplication script, and 29 by the attribution and sub-file runs, at a \head\ newer than the pre-registered clone; they are marked as such in the package. One sample repository with 1~GB of \head\ text exceeds the 400~MB snapshot cap of our byte attributor (the limit keeps the reference snapshot in memory) on 299 of 300 sampled commits and is excluded from the RQ2 medians. The repeat counter does not follow renames; following them changes the deleted-path share (31\% to 14\%) and version counts but not the corpus multiple (89.4$\times$ to 89.5$\times$).

\textbf{External validity.} RQ1--RQ3 are answered on the pre-registered sample, with the hand-picked, Python-heavy mature set as the upper tail. The frame excludes repositories under 200 commits; since the corpus multiple rises with commit count (tercile medians 14, 16 and 24$\times$), the excluded tail is likely lower. RQ4 concerns one benchmark and its 12 repositories, all Python; the mechanism, fixes spreading into every later version, is not specific to them. Cross-repository redundancy at GitHub scale is out of scope; a population-scale replication would go through World of Code or the Software Heritage graph~\cite{ma2021worldofcode,dicosmo2023softwareheritage}.

\section{Conclusion}
In a repository history, each file exists in many near-identical versions, a pangenome in the indexing sense, and the tools that deduplicate training data were built for collections of distinct files. Run across versions, the standard similarity-based deduplicator removes most of the repetition, taking a typical history from the equivalent of 19 copies of its current code to about two. Left to its default rule, it keeps a stale version of half of all files, and sorting newest first fixes that at almost no cost. And unless it happens to keep a version from before the fix, it keeps at least one copy of every benchmark fix the history contains, a fix that sits in a median of 18 later versions. Three rules follow for any corpus that is deduplicated: remove benchmark content across all versions, deduplicate across versions, and keep the newest. Edit corpora instead keep every version and charge training loss only on the change, and every pipeline should keep a count of versions per file. What the rules have in common is the unit they reason about: the lineage of a file over time, not the file.

\section*{Use of Generative AI}
The author designed the studies, drafted and reviewed the manuscript, and verified every reported number against the frozen outputs. AI coding assistants were used to write and run the analysis code and to improve the drafting of the text. All numbers are regenerated by the released scripts from the frozen outputs, and the author takes full responsibility for the content.

\section*{Data Availability}
The instruments, scripts, frozen outputs behind every table, pre-registrations, deviation logs, run metadata and the scripts that re-derive every reported number form a replication package (\pkgurl), released under the MIT License (code) and CC BY 4.0 (outputs and documents). The SWE-bench results are provided as per-repository and pooled aggregates.

\section*{Acknowledgment}
I thank Siva Padisetty, Manav Khurana and Nitin Singhal for their guidance and support on this work, and the GitLab Legal team for reviewing its release. I also thank Adam Hegyi, Ankit Bhatnagar, Ankit Panchal, John Mason, Ravi Kumar and Siddharth Dungarwal of the GitLab Code Search and GitLab Brain teams for helpful discussions and code reviews. The views expressed are the author's own.

\bibliographystyle{IEEEtran}
\bibliography{references}
\vspace{1ex}\noindent{\small\copyright{}2026 GitLab Inc.}
\end{document}